\documentclass[preprint,bibnotes]{revtex4}
\usepackage{amsfonts}
\usepackage{amsmath}
\usepackage{amssymb}
\usepackage{graphicx}

\begin{document}

\preprint{}
\title[\\
]{Different types of the Fulde-Ferrell-Larkin-Ovchinnikov states induced by
anisotropy effects}
\author{Dmitry Denisov}
\affiliation{Condensed Matter Theory Group, CPMOH, University of Bordeaux, 351 cours de
la Liberation, F-33405 Talence, France}
\author{Alexander Buzdin}
\altaffiliation{also at Institut Universitaire de France, Paris, France}
\affiliation{Condensed Matter Theory Group, CPMOH, University of Bordeaux, 351 cours de
la Liberation, F-33405 Talence, France}
\author{Hiroshi Shimahara}
\affiliation{Department of Quantum Matter Science, ADSM, Hiroshima University,
Higashi-Hiroshima 739-8530, Japan}

\begin{abstract}
The crystal structure determines both the Fermi surface and pairing symmetry
of the superconducting metals. It is demonstrated in the framework of the
general phenomenological approach that this is of the primary importance for
the determination of the structure of the Fulde-Ferrell-Larkin-Ovchinnikov
(FFLO) phase in the magnetic field. The FFLO modulation of the
superconducting order parameter may be revealed in the form of the higher
Landau level states or/and modulation along the magnetic field. The
transition between different FFLO\ states could occur with the temperature
variation or with the magnetic field rotation.
\end{abstract}

\volumeyear{}
\volumenumber{}
\issuenumber{}
\eid{}
\date{}
\received[Received text]{}
\revised[Revised text]{}
\accepted[Accepted text]{}
\published[Published text]{}
\startpage{1}
\endpage{}
\maketitle

\section{\protect\bigskip Introduction}

\bigskip \bigskip It is well known that in type II superconductors the
Abrikosov vortex state can be formed under a magnetic field. In most cases
the destruction of superconductivity happens due to the orbital effect.
However there can be a situation when paramagnetic effect plays an important
role in destruction of superconductivity (magnetic field acting only on
electron spins). In this case the non-uniform
Fulde-Ferrell-Larkin-Ovchinnikov (FFLO) state \cite{FuldeFerrel,
LarkinOvchinnikov} appears in superconductors, which is characterized by the
modulation of the order parameter. The structure of the FFLO phase in the
real compounds may be very rich \cite{GruenbergGunther, Bulaevskii74,
BuzdinBrison, BuzdinBrisonEuro, ShimaharaRainer97, Machida07,
MatsudaShimahara07, Shimahara98, ShimaharaMatsuo96, Klein2000}. Interplay of
orbital and paramagnetic effects has been described in the isotropic model
by Gruenberg and Gunther in \cite{GruenbergGunther}, where they calculated
critical field and structure of the order parameter. It was found in \cite%
{GruenbergGunther} that the orbital effect is detrimental to the FFLO state,
but still such a state can exist if the ratio of pure orbital effect $%
H_{c2}^{orb}(0)$ and pure paramagnetic limit $H_{p}(0)$ is larger than 1.28,
i.e. the Maki parameter $\alpha _{M}=\sqrt{2}H_{c2}^{orb}(0)/H_{p}(0)$ must
be larger 1.8. Pure paramagnetic limit at $T=0$ can be estimated as $%
H_{p}(0)=\Delta _{0}/\sqrt{2}\mu _{B}$, where $\Delta _{0}$ is BCS gap at $%
T=0$ and $\mu _{B}$ is the Bohr magneton. In \cite{GruenbergGunther} the
modulation was studied using a zero Landau level function, which holds true
only for moderate Maki parameter $\alpha _{M}<9$. It was found in \cite%
{BuzdinBrison} that for large values of Maki parameter $\alpha _{M}>9$ the
higher Landau level solutions become relevant. In this case the critical
field $H_{c2}(T)$ consists of several curves each corresponding to a
different Landau level solution. The analysis of the orbital effect in the
FFLO state \cite{GruenbergGunther, BuzdinBrison} were performed for the
isotropic metals with $s$-wave type of pairing. However in \cite{Brison95}
it was demonstrated that it readily generalized for the case of the metals
with elliptic Fermi surface. \bigskip In such a case the Maki parameter
becomes angular dependent. For example for the case of the quasi-2D or
anisotropic 3D superconductors $\alpha _{M}$ increases dramatically for the
in-plane field orientation. Therefore we may expect the transitions between
the usual FFLO state with zero Landau levels \cite{GruenbergGunther} to the
state with higher Landau levels \cite{BuzdinBrison, BuzdinBrisonEuro,
Bulaevskii74, ShimaharaRainer97} when the magnetic field is tilted from the
perpendicular orientation to the parallel one. Also crossover from the pure
FFLO state to the vortex states with higher Landau levels indexes in the
model of quasi-2D system has been predicted in \cite{ShimaharaBook}. In real
compounds the deviation of the Fermi surface from the elliptic form is
crucial for the adequate description of the FFLO state as well as the type
of the superconductivity pairing (e.g. $s$- or $d$-wave) \cite{Mineev06,
BuzdinMatsuda07, Konschelle07}. This circumstance is related with a fact
that the description of the FFLO state in the framework of Ginzburg-Landau
approach needs the consideration of the higher-order derivatives of the
order parameter in addition to the usual gradient terms. For example, in the
case of pure paramagnetic effect the critical field and modulation vector $q$
strongly depend on anisotropy or nesting properties of the Fermi surface 
\cite{Shimahara96, Shimahara99, Shimahara2000, Shimahara02}. In this article
we consider a realistic case with a non-elliptic Fermi surface and for the
definiteness we restrict ourself to the tetragonal symmetry. For example
quasi-two-dimensional superconductor CeCoIn$_{5\text{ }}$\cite%
{MatsudaShimahara07} provides favorable conditions for the formation of the
FFLO state and it has a tetragonal symmetry.

\bigskip A characteristic feature of the FFLO state is the existence of a
tricritical point (TCP) in the field-temperature phase diagram \cite%
{SaintJamesSarma}. TCP is the meeting point of three transition lines
separating the normal metal, the uniform superconductivity and the FFLO
state. Formation of the FFLO state near the TCP may be described by modified
Ginzburg-Landau functional (MGL) \cite{BuzdinKachKachi97}. Appearance of the
non-uniform state is related with a change of the sign of the coefficient $g$
at the gradient term $g\left\vert \Pi _{i}\Psi \right\vert ^{2}$ in the free
energy density. In the standard Ginzburg-Landau theory the coefficient $g$
is positive. Here it vanishes at the TCP ($T^{\ast },H_{c2}(T^{\ast })$),
and then becomes negative for $T<T^{\ast }$. The absolute value of $g$ grows
as we move further from the TCP, for example with increasing of the magnetic
field or lowering temperature. A negative $g$ means that the modulated state
has a lower free energy than the uniform one. In order to obtain the
modulation vector one needs to include the term with higher order
derivatives in the MGL functional \cite{BuzdinKachKachi97}.

In this paper we study the effects of crystal (or pairing) anisotropy on the
FFLO phase. Using MGL approach we introduce free energy density $\mathcal{F}$
describing tetragonal system. We examine the case of the Fermi surface close
to elliptic one. Therefore $\mathcal{F}$ can be divided into isotropic and
perturbative parts. We demonstrate that the higher Landau level solutions
may be realized for arbitrary values of Maki parameter in contrast with
isotropic model. This is a special mechanism of the higher Landau level
phase formation in 3D system. Moreover depending on various type of
deviation of the Fermi surface from isotropic form three possible solutions
for the FFLO state can be realized: (a) maximum modulation occurs along the
magnetic field with zero Landau level state, (b) both modulation and higher
Landau level state, (c) highest possible Landau level and no modulation
along the field (or modulation with very small wave-vector). Moreover due to
the specific form of the Fermi surface the variation of magnetic field
orientation may provoke transitions between the states with different Landau
levels.

The main goal of the present paper is to demonstrate that in the presence of
the orbital effect and for the realistic Fermi surface the very different
types of the FFLO state could be realized. In particular, if the preferred
modulation direction is perpendicular to the magnetic field this can results
in the formation of the higher Landau levels mixed state with no modulation
along the field at all. Our approach is fully justified near the TCP\ and
for superconductors with large Maki parameters. However qualitatively it
provides the understanding of the FFLO\ state at all temperatures and for
arbitrary strength of the orbital effect. Here we calculate the line of the
second order transition from the normal to the superconducting state. For
this purpose we use the quadratic over the superconducting order parameter
MGL functional. To describe the properties of the FFLO state it is needed to
retain the higher order terms over the superconducting order parameter. The
situation is completely analogous with that of the Abrikosov vortex lattice.
From the symmetry reasons it is clear that the transitions between the mixed
states describing by the different Landau levels will be the first order
transitions. However the appearance of the modulation along the magnetic
field may occur through a continuous transition. All these interesting
questions deserve further studies but they are well beyond the scope of the
present article.

\bigskip

\section{FFLO state in anisotropic Ginzburg-Landau model}

The most general form of the MGL functional quadratic over $\Psi $ is%
\begin{equation}
\mathcal{F}=\alpha \left\vert \Psi \right\vert
^{2}-\sum_{i=1}^{3}g_{i}\left\vert \Pi _{i}\Psi \right\vert
^{2}+\sum_{i=1}^{3}\gamma _{i}\left\vert \Pi _{i}^{2}\Psi \right\vert
^{2}+\sum\limits_{i\neq j}\varepsilon _{ij}\left\vert \Pi _{i}\Pi _{j}\Psi
\right\vert ^{2},  \label{eq:Fgeneral}
\end{equation}%
where $\alpha (H,T)=\alpha _{0}(T-T_{cu}(H)),$ $T_{cu}(H)$ is transition
temperature into the uniform superconducting state, $\Pi _{i}=-i\hbar \frac{%
\partial }{\partial x_{i}}-\frac{2e}{c}A_{i}$ are momentum operators and $%
A_{i}$ are the components of the vector potential, further we put $\hbar =1$%
. Assuming the tetragonal symmetry in the case of the pure elliptic Fermi
surface we have following relation for the effective \ mass $m_{z}\neq
m_{x}=m_{y}=m$. The elliptical Fermi surface could be transformed into
isotropic one by the following scaling transformation $z^{\prime }=\sqrt{%
m_{z}/m}z$ \cite{Brison95}. Components of the magnetic field also transform
as $H\longrightarrow H^{\prime }=(\frac{m}{m_{z}}H_{x},\frac{m}{m_{z}}%
H_{y},H_{z})$. Further we suppose that actual Fermi surface deviation from
the elliptical form is small and after corresponding scaling transformation
the functional (\ref{eq:Fgeneral}) for the tetragonal symmetry is written as

\begin{gather}
\mathcal{F}=\alpha \left\vert \Psi \right\vert
^{2}-g\sum_{i=1}^{3}\left\vert \Pi _{i}\Psi \right\vert ^{2}+\gamma
\left\vert \sum_{i=1}^{3}\Pi _{i}^{2}\Psi \right\vert ^{2}+\varepsilon
_{z}\left\vert \Pi _{z}^{2}\Psi \right\vert ^{2}+\frac{\varepsilon _{x}}{2}%
\left( \left\vert \Pi _{x}\Pi _{y}\Psi \right\vert ^{2}+\left\vert \Pi
_{y}\Pi _{x}\Psi \right\vert ^{2}\right)  \label{eq:Ftetra} \\
+\frac{\widetilde{\varepsilon }}{2}\left( \left\vert \Pi _{z}\Pi _{x}\Psi
\right\vert ^{2}+\left\vert \Pi _{x}\Pi _{z}\Psi \right\vert ^{2}+\left\vert
\Pi _{z}\Pi _{y}\Psi \right\vert ^{2}+\left\vert \Pi _{y}\Pi _{z}\Psi
\right\vert ^{2}\right) .  \notag
\end{gather}%
Coefficients $g,\gamma ,\varepsilon _{z},\varepsilon _{x},\tilde{\varepsilon}
$ depend on structure of the Fermi surface, but for the FFLO\ appearance $g$
must be positive. The terms $-g\sum_{i=1}^{3}\left\vert \Pi _{i}\Psi
\right\vert ^{2}+\gamma \left\vert \sum_{i=1}^{3}\Pi _{i}^{2}\Psi
\right\vert ^{2}$ describe the elliptic form of the Fermi surface (for $s$%
-wave superconductivity) and terms with coefficients $\varepsilon _{z}$, $%
\varepsilon _{x}$ and $\widetilde{\varepsilon }$ \bigskip are considered as
perturbation.

\begin{figure}[tbp]
\begin{center}
\includegraphics[height=3in]{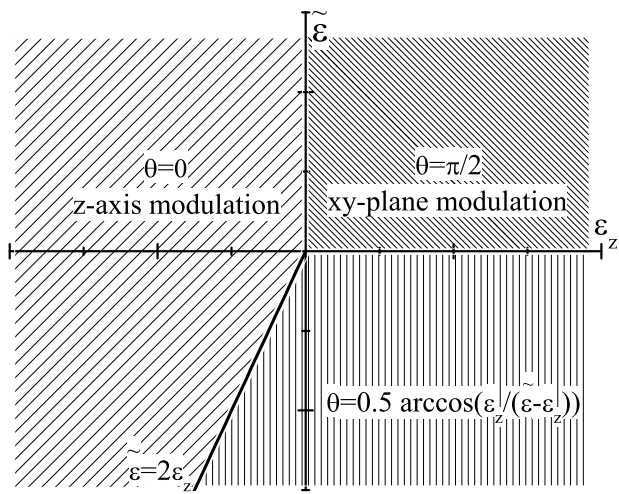}
\end{center}
\caption{Modulation ($\widetilde{\protect\varepsilon },\protect\varepsilon %
_{z}$) diagram in the case of the absence of the orbital effect (pure
paramagnetic limit). Areas with different patterns correspond to different
orientation of the wave-vector modulation. The phase diagram does not depend
on the $\protect\varepsilon _{x}$ value.}
\label{fig:noH}
\end{figure}

Without orbital effect the momentum operators are simplified to $\Pi _{i}=-i%
\frac{\partial }{\partial x_{i}}$ and the solution for the order parameter
could be presented as $\Psi =\Psi _{q}\exp (i\vec{q}\cdot \vec{r})$. In the
case of elliptic Fermi surface we have a degeneracy over direction of the
FFLO modulation $q$. The crystal structure effects are expressed via the
terms with $\varepsilon _{z}$, $\varepsilon _{x}$,$\ \widetilde{\varepsilon }
$ and they lift this degeneracy and determine the direction of the FFLO
modulation. The free energy density is written as%
\begin{equation}
\mathcal{F}=\sum_{q}\left\{ \alpha -gq^{2}+\gamma q^{4}+\varepsilon
_{z}q^{4}\cos ^{4}\theta +\varepsilon _{x}q^{4}\sin ^{4}\theta \cos
^{2}\varphi \sin ^{2}\varphi +\widetilde{\varepsilon }q^{4}\sin ^{2}\theta
\cos ^{2}\theta \right\} \left\vert \Psi _{q}\right\vert ^{2},
\label{eq:Fnoorb}
\end{equation}%
\bigskip where we have used spherical system for $\vec{q}$: $q_{x}=q\cos
\varphi \sin \theta $, $q_{y}=q\sin \varphi \sin \theta $, $q_{z}=q\cos
\theta $. Due to the tetragonal symmetry the modulation in $xy$-plane is
either parallel to $x$ or $y$ axis ($\varepsilon _{x}>0$) or along the
bisector ($\varepsilon _{x}<0$). For definiteness we may suppose that $%
\varepsilon _{x}>0$. Note that in the case $\varepsilon _{x}<0$ the rotation
of the $xy$ axis by $\pi /4$ provides us the same functional (\ref{eq:Fnoorb}%
) with renormalized coefficients $\varepsilon _{z}^{\prime }$ and $%
\widetilde{\varepsilon }^{\prime }$ but with $\varepsilon _{x}>0$. The wave
vector of modulation $q$ will be in $xy$-plane if $\widetilde{\varepsilon }%
>2\varepsilon _{z}$, $\varepsilon _{z}<0$, parallel to $z$ axis if $%
\widetilde{\varepsilon }>0$, $\varepsilon _{z}>0$ (see Fig.~\ref{fig:noH}).
For the region $\widetilde{\varepsilon }<0,\widetilde{\varepsilon }%
<2\varepsilon _{z}$ the direction of modulation is at angle $\theta =\frac{1%
}{2}\arccos \frac{\varepsilon _{z}}{(\widetilde{\varepsilon }-\varepsilon
_{z})}$ to the $xy$-plane. Therefore the crystal anisotropy (or/and pairing
anisotropy) lifts the degeneracy over the direction of the FFLO\ modulation
and the whole diagram in the ($\varepsilon _{z}$, $\widetilde{\varepsilon }$%
) plane is presented in Fig.~\ref{fig:noH}.

\section{\protect\bigskip Orbital effect for the magnetic field applied
along z axis}

\bigskip The exact solution of the linearized equation for $\Psi (\vec{r})$
is unavailable in general case if the orbital effect is taken into account.
The Landau level solution with additional modulation along the field works
only for the case of elliptical Fermi surface. In this case we obtain the
degeneracy over modulation $q$ and Landau level $n$, when we move from the
TCP. However if the anisotropy effects are taken into account they lift this
degeneracy. We demonstrate that depending on the parameters of the system
very different types of the FFLO state could be realized. \bigskip

We begin with the case when magnetic field $H$ is applied along tetragonal $%
z $ axis and the gauge is chosen as $A$=$(yH,0,0)$. Following for example
Ref.~\cite{ShimaharaRainer97}\ we can express our operators $\Pi _{i}$ using
boson operators of creation $\eta $\ and annihilation $\eta ^{+}$ as $\Pi
_{x}=i\frac{1}{\sqrt{2}\xi _{H}}\left( \eta -\eta ^{+}\right) $, $\Pi _{y}=%
\frac{1}{\sqrt{2}\xi _{H}}\left( \eta +\eta ^{+}\right) $, where $\xi _{H}=%
\sqrt{\frac{c}{2eH}}$. Our goal is to find $T_{c}(H)$, which is the
transition temperature into the FFLO state, i.e. we need to find the
solution which gives the maximum of $\alpha (H,T)=\alpha
_{0}(T_{c}-T_{cu}(H))$. To do this we use the variation method \cite%
{AbriskovBook88} and look for a maximum of $\alpha $ written as

\begin{equation}
\alpha (H,T)=\max \left\{ \frac{\int \left[ \alpha \left\vert \Psi
\right\vert ^{2}-\mathcal{F}\right] d^{3}r}{\int \left\vert \Psi \right\vert
^{2}d^{3}r}\right\} .
\end{equation}

In general case $\Psi (x,y,z)=\sum C_{n}\varphi _{n}(x,y)e^{iq_{z}z}$, but
since the anisotropy effects are small we can approximate our solution only
with a single Landau level function $\varphi _{N}$ \cite{LandauFunctions},
which is well known solution for the system with isotropic form of the Fermi
surface. In our calculations we use the following properties of Landau
functions: $\int \varphi _{n}\varphi _{m}d^{3}r=\delta _{nm}$, $\eta \varphi
_{n}=\sqrt{n}\varphi _{n-1}$, $\eta ^{+}\varphi _{n}=\sqrt{n+1}\varphi
_{n+1} $; we also normalize $C_{n}$ so that $\int \left\vert \Psi
\right\vert ^{2}d^{3}r=1$. To neglect the other Landau level functions ($%
n\neq N$) in our $\Psi $ representation their corresponding coefficients
should be small comparing to the coefficient $C_{N}$.\bigskip\ This leads to
the following condition for the applicability of the single level
approximation $\xi _{H}^{-2}\gg \frac{g}{\gamma }\sqrt{\frac{\varepsilon }{%
\gamma }}$ which is determined by the Fermi surface deviation $\varepsilon $
from the elliptic form ($\varepsilon $ is of the order of average between $%
\widetilde{\varepsilon }$, $\varepsilon _{z}$ and $\varepsilon _{x}$ and its
exact value depends on magnetic field orientation). Calculating $\alpha
(H,T) $ using the operators $\eta $\ and $\eta ^{+}$ we can express it in
terms of $q_{z}$ and $\xi _{n}^{-2}=\xi _{H}^{-2}(2n+1)$ as

\begin{gather}
\alpha (H,T)=\max \left\{ g\left[ q_{z}^{2}+\xi _{n}^{-2}\right] -\gamma %
\left[ q_{z}^{2}+\xi _{n}^{-2}\right] ^{2}\right.  \label{eq:Fz2z} \\
\left. -\varepsilon _{z}q_{z}^{4}-\varepsilon _{x}\frac{1}{8}\xi _{n}^{-4}-%
\tilde{\varepsilon}q_{z}^{2}\xi _{n}^{-2}-\varepsilon _{x}\frac{5}{8}\xi
_{H}^{-4}\right\} .  \notag
\end{gather}

\begin{figure}[tbp]
\begin{center}
\includegraphics[height=3in]{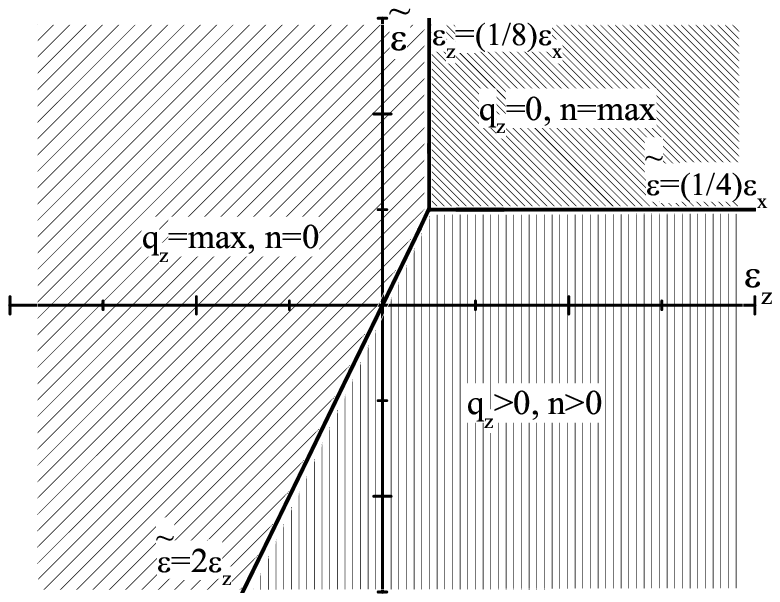}
\end{center}
\caption{Modulation diagram in the case when the magnetic field applied
along $z$ axis. There are 3 areas on the diagram corresponding to 3 types of
the solution for modulation vector $q_{z}$ and Landau level $n$. Modulation
direction is always parallel to the applied field and $\protect\varepsilon %
_{x}$ here is chosen equal to $\protect\gamma $.}
\label{fig:z}
\end{figure}

\bigskip If we consider only unperturbed part $g\left[ q_{z}^{2}+\xi
_{n}^{-2}\right] -\gamma \left[ q_{z}^{2}+\xi _{n}^{-2}\right] ^{2}$, put $%
u=q_{z}^{2}+\xi _{n}^{-2}$ and take derivative with respect to $u$ we obtain
the maximum point at $u_{0}=g/2\gamma $. When the system is close to the
TCP, $g$ is small so that $g/2\gamma <\xi _{H}^{-2}$. In this case we obtain
the lowest Landau level $n=0$ and no modulation along $z$ ($q_{z}=0$). If we
move further from the TCP and $g$ grows $\left( \xi _{H}^{-2}<g/2\gamma
<3\xi _{H}^{-2}\right) $ then $n$ remains equal to $0$, but $q_{z}$ is not.
If Maki parameter is large then our approach will still be valid even far
away from the TCP. In this case $g/2\gamma \geq 3\xi _{H}^{-2}$ and $n$ can
be larger than 0. Hence, we have degeneracy over choosing of Landau level $n$
and $q_{z}$. But as was written earlier taking into account small
perturbative terms with $\varepsilon _{z},\varepsilon _{x},\tilde{\varepsilon%
}$ we remove this degeneracy and find the maximum $\alpha (H,T)$ with a
respect to $q_{z}=\sqrt{u_{0}-\xi _{n}^{-2}}$. There are three possible
types of solutions (combinations of $n$ and $q_{z}$) depending on $%
\varepsilon _{z},\varepsilon _{x},\tilde{\varepsilon}$: (a) maximum
modulation $q_{z}=\sqrt{g/2\gamma -\xi _{H}^{-2}}$ and zero Landau level $n$%
; (b) non zero modulation $q_{z}=\sqrt{\frac{g}{2\gamma }\left( \varepsilon
_{x}-4\tilde{\varepsilon}\right) /\left( 8\varepsilon _{z}+\varepsilon _{x}-8%
\tilde{\varepsilon}\right) }$ with $n=\left[ \frac{1}{2}\left( \xi
_{H}^{2}\left( u_{0}-q_{z}^{2}\right) -1\right) \right] ,$where brackets $[$ 
$]$ mean that only integer part is taken; (c) highest possible Landau level $%
n=\left[ \frac{g}{4\gamma }\xi _{H}^{2}-\frac{1}{2}\right] $ and near zero
modulation. All these cases are shown in Fig.~\ref{fig:z}. However due to
integer nature of $n$ the modulation $q_{z}$ has a very special behavior in
cases b) and c), it changes abruptly every time when our solution jumps from
one to another Landau level. It should be noted that in the case c) the
wave-vector of modulation $q_{z}$ instead of being zero oscillates with $H$
(or with $g$, when we move further form the TCP) due to the mismatch of $%
u_{0}=g/2\gamma $ and $\xi _{n}^{-2}=\xi _{H}^{-2}(2n+1)$. The diagram shown
in Fig.~\ref{fig:z} looks similar to that in Fig.~\ref{fig:noH} with the
exception of the shift along both axis due to the presence of the $%
\varepsilon _{x}$ coefficient. When we get zero wave-vector of modulation in
Fig.~\ref{fig:z} the same area in Fig.~\ref{fig:noH} corresponds to
modulation in $xy$-plane. If modulation vector $q$ along the applied
magnetic field is zero then modulation could arise in the direction
perpendicular to the field. When $n$ and $q_{z}$ is intermediate (not
maximum and not zero, lower right part of the diagrams) FFLO modulation
could be formed in both perpendicular and parallel directions to the field ($%
q_{\parallel }^{2}+q_{\perp }^{2}=g/2\gamma -\xi _{H}^{-2}$). It should be
noted that we cannot achieve a smooth transition from one diagram to another
by decreasing $H$ to 0 due the condition of single Landau level
approximation $\xi _{H}^{-2}\gg \frac{g}{\gamma }\sqrt{\frac{\varepsilon }{%
\gamma }}$, despite the fact that the only difference between diagrams in
Figs.~\ref{fig:noH} and \ref{fig:z} is shift of the intersection point due
to the presence of the $\varepsilon _{x}$ coefficient. The same is true for
the case $H\parallel x\ $where difference will be more significant.

\begin{figure}[tbp]
\begin{center}
\includegraphics[height=3in]{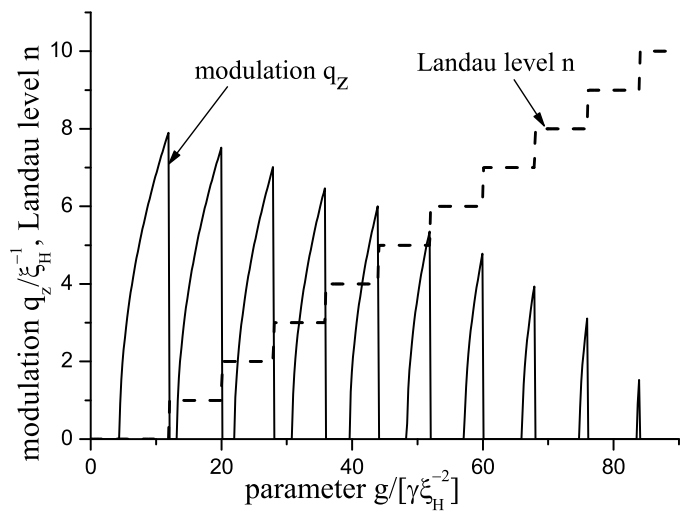}
\end{center}
\caption{Dependence of residual modulation on the parameter $g$ (normalized
to $\protect\gamma \protect\xi _{H}^{-2}$). The wave vector of modulation $q$
is shown here by the solid line and measured in units of $\protect\xi %
_{H}^{-1}$ . The values of Landau level $n$ is shown by the dashed line. The
parameter $\tilde{\protect\varepsilon}$ is chosen here as $0.2\protect\gamma$%
.}
\label{fig:residual}
\end{figure}

\textbf{Maximum Landau level and residual modulation}. \bigskip Due to the
fact that $n$ is integer the wave-vector of modulation $q_{z}$ may not be
equal exactly to zero but to some value less than $\sqrt{u_{0}-\xi _{n}^{-2}}
$ when $n$ is maximum. To calculate this value we maximize $\alpha (H,T)$
from the Eq.~(\ref{eq:Fz2z}) again but this time $\xi _{n}^{-2}=\xi
_{H}^{-2}(2n+1)$ will be treated as constant. We obtain $q_{z}^{2}=\max
\left( 0,\frac{u_{0}-\xi _{n}^{-2}-(\tilde{\varepsilon}/2\gamma )\xi
_{n}^{-2}}{1+(\varepsilon _{z}/\gamma )}\right) $ and the general solution
for residual modulation $q_{z}$ will oscillate with $H$ or with $g$
(absolute value of $g$ is increasing when we are moving away from the TCP).
Wave vector of modulation is zero when $\xi _{n}^{-2}$ is close to $%
u_{0}=g/2\gamma $, then is start to increase linearly with $g$\ until it
drops to zero again when the solution "jumps" to another Landau level (Fig.~%
\ref{fig:residual}). With the increasing effect of anisotropy, the area of
zero wave-vector modulation widens, and in the limiting case it will cover
all parameter range. Similar results were obtained for isotropic case at $%
\alpha _{M}>9$ at low temperature \cite{BuzdinBrison}.

\section{\protect\bigskip The case of the magnetic field applied along x axis%
}

\begin{figure}[tbp]
\begin{center}
\includegraphics[height=3in]{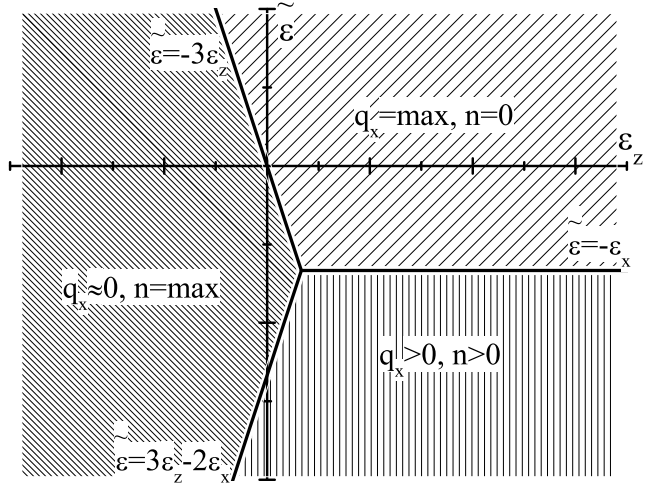}
\end{center}
\caption{Modulation diagram ($\widetilde{\protect\varepsilon },\protect%
\varepsilon _{z}$) in the case when the magnetic field is applied along $x$
axis. There are three areas on the diagram corresponding to different types
of the solution for modulation vector $q_{x}$ and Landau level $n$.
Modulation direction is always parallel to the applied field. The position
of the intersection point is determined by the coefficient $\protect%
\varepsilon _{x}$.}
\label{fig:Hx}
\end{figure}

For magnetic field applied along the $x$ axis we choose $\vec{A}$ as $%
(0,0,yH)$, the order parameter $\Psi (x,y,z)=\varphi _{N}(y)e^{iq_{x}x}$
where $q_{x}$ is modulation along the $x$ axis, and reintroduce creation and
annihilation operators as $\eta =\frac{\xi _{H}}{\sqrt{2}}\left( \Pi
_{y}-i\Pi _{z}\right) $ and $\eta ^{+}=\frac{\xi _{H}}{\sqrt{2}}\left( \Pi
_{y}+i\Pi _{z}\right) $. Repeating the same calculations as for the $H\Vert
z $ case we have

\begin{gather}
\alpha (H,T)=\max \left\{ g\left[ q_{x}^{2}+\xi _{n}^{-2}\right] -\gamma %
\left[ q_{x}^{2}+\xi _{n}^{-2}\right] ^{2}\right.  \\
\left. -\varepsilon _{z}\frac{3}{8}\xi _{n}^{-4}-\frac{1}{2}\varepsilon
_{x}q_{x}^{2}\xi _{n}^{-2}-\frac{1}{2}\tilde{\varepsilon}q_{x}^{2}\xi
_{n}^{-2}-\tilde{\varepsilon}\frac{1}{8}\xi _{n}^{-4}-\varepsilon _{z}\frac{3%
}{8}\xi _{H}^{-4}-\tilde{\varepsilon}\frac{5}{8}\xi _{H}^{-4}\right\} . 
\notag
\end{gather}%
\ Again we put $u=q_{x}^{2}+\xi _{n}^{-2}$ and find that maximum point $%
u_{0}=g/2\gamma $ is the same for unperturbed part. The degeneracies over $%
q_{x}$ and $n$ are removed in a similar way by taking into account the
perturbative terms $\varepsilon _{z},\varepsilon _{x},\tilde{\varepsilon}$.
Diagram for maximums of $\alpha (H,T)$ in the case with magnetic field
applied along $x$ axis is shown in Fig.~\ref{fig:Hx}. Three main areas of
the diagrams are similar to the ones shown in Fig.~\ref{fig:z}: (a) maximum
modulation $q_{x}=\sqrt{g/2\gamma -\xi _{H}^{-2}}$ and zero Landau level $n$%
; (b) non zero modulation $q_{x}=\sqrt{\frac{g}{2\gamma }\left( 3\varepsilon
_{z}-\tilde{\varepsilon}-2\varepsilon _{x}\right) /\left( 3\varepsilon _{z}-3%
\tilde{\varepsilon}-4\varepsilon _{x}\right) }$ with $n=\left[ \frac{1}{2}%
\left( \xi _{H}^{2}\left( u_{0}-q_{x}^{2}\right) -1\right) \right] $; (c)
highest possible Landau level $n=\left[ \frac{g}{4\gamma }\xi _{H}^{2}-\frac{%
1}{2}\right] $ and residual modulation described earlier. Diagrams shown in
Figs.~\ref{fig:Hx} and \ref{fig:noH} have some similarities with a respect
to the conditions for the different types of solutions. On both diagrams the
FFLO modulation along $x$ (or $y$) axis corresponds to the upper-right
quarter and there is no modulation in $xy$ plane in the left quarters of the
diagrams.

\section{\protect\bigskip \protect\bigskip Magnetic field applied in xy plane%
}

\bigskip If a magnetic field $H$ is applied in the $xy$-plane ($\beta $ is
the angle between $\vec{H}$ and $x$ axis), then it is convenient to rotate $%
x $, $y$ axis around $z$ by angle $\beta $ to reduce the problem to the case 
$H\parallel x$. Under this rotation the terms with coefficients $g$, $\gamma 
$, $\varepsilon _{z}$ remain unchanged and the rest parts are transformed
according to rules $x^{\prime }=x\cos \beta +y\sin \beta ,$ $y^{\prime
}=y\cos \beta -x\sin \beta $

\begin{eqnarray}
\Pi _{x} &=&\Pi _{x}^{\prime }\cos \beta -\Pi _{y}^{\prime }\sin \beta \\
\Pi _{y} &=&\Pi _{x}^{\prime }\sin \beta +\Pi _{y}^{\prime }\cos \beta \\
\Pi _{z} &=&\Pi _{z}^{\prime }=-i\frac{\partial }{\partial z}-\frac{2e}{c}%
y^{\prime }H
\end{eqnarray}

\begin{figure}[tbp]
\begin{center}
\includegraphics[height=3in]{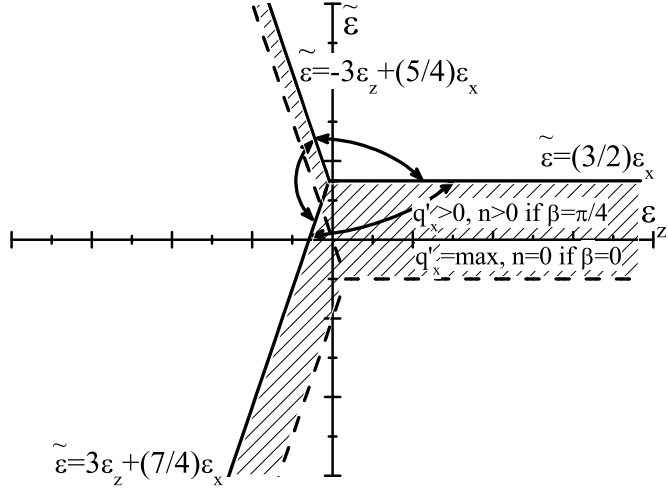}
\end{center}
\caption{Modulation diagram in the case when the magnetic field applied in
xy-plane. Solid separation lines correspond to the case of $\protect\beta =%
\protect\pi /4$ and dashed lines to the case of $\protect\beta =0$.
Patterned area corresponds to the region where the type of the solution for $%
q$ and $n$ could be changed during rotation along $z$ axis.}
\label{fig:Hxy}
\end{figure}

The operators $\eta $\ and $\eta ^{+}$ are expressed using new $\Pi
_{x}^{\prime }$, $\Pi _{x}^{\prime }$ and $\Pi _{z}^{\prime }$ as before in
the case $H\parallel x$. Due to the symmetry of the problem only $%
\varepsilon _{x}$ term in Eq.~(\ref{eq:Ftetra}) acquires the dependence on $%
\beta $ in the final expression for $\alpha (H,T)$

\begin{gather}
\alpha (H,T)=\max \left\{ g\left[ q_{x}^{\prime 2}+\xi _{n}^{-2}\right]
-\gamma \left[ q_{x}^{\prime 2}+\xi _{n}^{-2}\right] ^{2}\right.  \notag \\
\left. -\varepsilon _{z}\frac{3}{8}\xi _{n}^{-4}-\frac{1}{2}\varepsilon
_{x}q_{x}^{\prime 2}\xi _{n}^{-2}-\varepsilon _{x}\frac{\sin ^{2}2\beta }{4}%
\left( q_{x}^{\prime 4}-3q_{x}^{\prime 2}\xi _{n}^{-2}+\frac{3}{8}\xi
_{n}^{-4}+\frac{3}{8}\xi _{H}^{-4}\right) \right. \\
\left. -\frac{1}{2}\tilde{\varepsilon}q_{x}^{\prime 2}\xi _{n}^{-2}-\tilde{%
\varepsilon}\frac{1}{8}\xi _{n}^{-4}-\varepsilon _{z}\frac{3}{8}\xi
_{H}^{-4}-\tilde{\varepsilon}\frac{5}{8}\xi _{H}^{-4}\right\} ,  \notag
\end{gather}%
where $q_{x}^{\prime }$ is modulation along the new $x^{\prime }$ axis
parallel to the magnetic field $H$. Directly from the $\varepsilon _{x}$
term it can be concluded that the angles $\beta =0,\pm \frac{\pi }{2},\pi $
will lead to the old results, when magnetic field is applied along $x$ or $y$
axis. The main results for the maximum of $\alpha (H,T)$ will be similar to
the case $H\parallel x$, with the only exception that separation lines on
phase diagram are changed to three lines: $\tilde{\varepsilon}=-3\varepsilon
_{z}+\frac{5}{4}\sin ^{2}2\beta \varepsilon _{x},$ $\tilde{\varepsilon}%
=3\varepsilon _{z}+\left( \frac{15}{4}\sin ^{2}2\beta -2\right) \varepsilon
_{x}$ and $\tilde{\varepsilon}=\left( \frac{5}{2}\sin ^{2}2\beta -1\right)
\varepsilon _{x}$. However this change only affects the initial shift of the
diagram from the center. For example in Fig.~\ref{fig:Hxy} the case $\beta =%
\frac{\pi }{2}(n+\frac{1}{2})$ is shown and the intersection point has
shifted to the opposite quarter of the graph. \bigskip For general values of 
$\beta $, the intersection point is situated at $\left( \varepsilon _{z},%
\tilde{\varepsilon}\right) $ = $\left( \left( -\frac{5}{12}\sin ^{2}2\beta +%
\frac{1}{3}\right) \varepsilon _{x},\left( \frac{5}{2}\sin ^{2}2\beta
-1\right) \varepsilon _{x}\right) $ on a line segment connecting the two
intersection points for $\beta =\frac{\pi }{2}n$ and $\beta =\frac{\pi }{2}%
(n+\frac{1}{2})$. On phase diagram like in previous cases we have 3 possible
solutions:

\begin{enumerate}
\item $q_{x}^{\prime 2}=u_{0}\left( 3\varepsilon _{z}-\tilde{\varepsilon}%
-2\varepsilon _{x}+\frac{15}{4}\sin ^{2}2\beta \text{ }\varepsilon
_{x}\right) /\left( 3\varepsilon _{z}-3\tilde{\varepsilon}-4\varepsilon _{x}+%
\frac{35}{4}\sin ^{2}2\beta \text{ }\varepsilon _{x}\right) ,$ $n=\left[ 
\frac{1}{2}\left( \xi _{H}^{2}\left( u_{0}-q_{x}^{\prime 2}\right) -1\right) %
\right] $, when $\tilde{\varepsilon}<\min \left( 3\varepsilon
_{z}-2\varepsilon _{x}+\frac{15}{4}\sin ^{2}2\beta \varepsilon _{x},\text{ }%
\varepsilon _{x}(\frac{5}{2}\sin ^{2}2\beta \varepsilon _{x}-1)\right) $.

\item $q_{x}^{\prime }\approx 0$ (near zero residual modulation), maximum $n$
in this case will be equal to $\left[ \frac{g}{4d}\xi _{H}^{2}-\frac{1}{2}%
\right] $. This corresponds to $-3\varepsilon _{z}+\frac{5}{4}\sin
^{2}2\beta \varepsilon _{x}>\tilde{\varepsilon}>3\varepsilon
_{z}-2\varepsilon _{x}+\frac{15}{4}\sin ^{2}2\beta \varepsilon _{x}$.

\item $q_{x}^{\prime }=\sqrt{g/2\gamma -\xi _{H}^{-2}}$ - maximum modulation
along the magnetic field with $n=0$. This corresponds to $\tilde{\varepsilon}%
>\max \left( \varepsilon _{x}(\frac{5}{2}\sin ^{2}2\beta \varepsilon _{x}-1),%
\text{ }-3\varepsilon _{z}+\frac{5}{4}\sin ^{2}2\beta \varepsilon
_{x}\right) $.
\end{enumerate}

\begin{figure}[tbp]
\begin{center}
\includegraphics[height=3in]{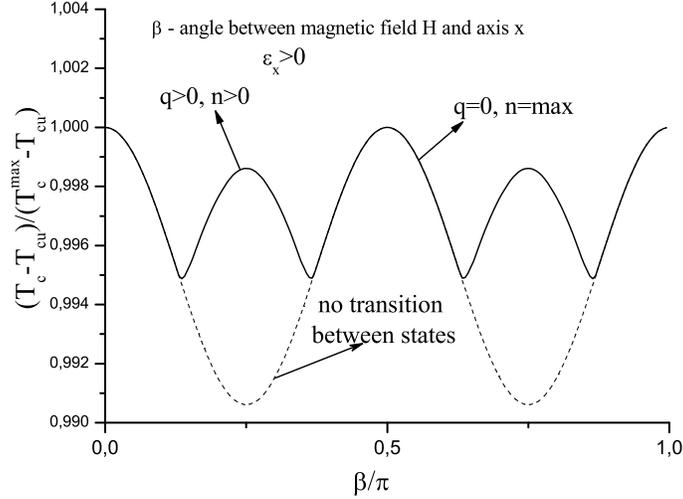}
\end{center}
\caption{Transition temperature dependence on the angle $\protect\beta $ of
the magnetic field in $xy$-plane. For illustration we have chosen $\protect%
\varepsilon _{z}=\tilde{\protect\varepsilon}=0,$ $\protect\varepsilon %
_{x}=0.1\protect\gamma ,$ $g$ (normalized to $\protect\gamma \protect\xi %
_{H}^{-2}$) equal to 100. There are switching between two types of the
solution: ($q>0,n>0$) and ($q=0,n=\max $).}
\label{fig:Tcp}
\end{figure}

\begin{figure}[tbp]
\begin{center}
\includegraphics[height=3in]{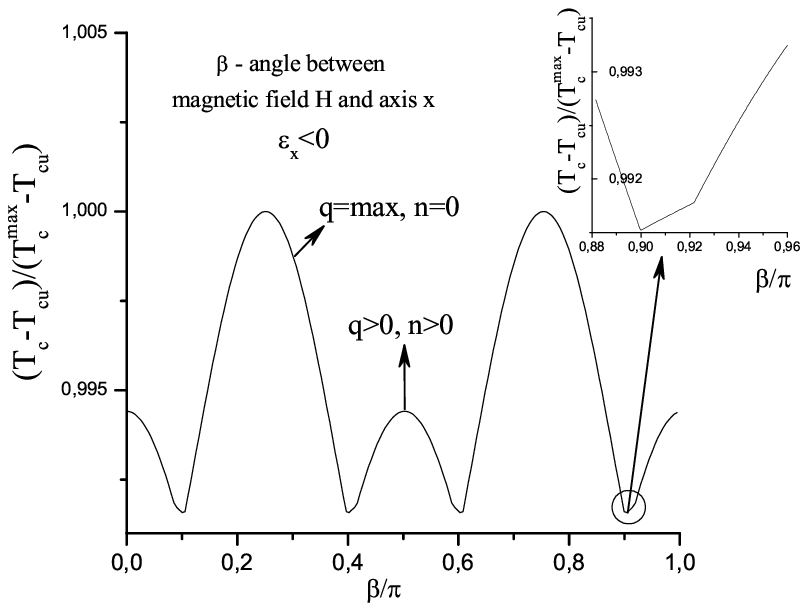}
\end{center}
\caption{Transition temperature dependence on the angle $\protect\beta $ of
the magnetic field in $xy$-plane. For illustration we have chosen $\protect%
\varepsilon _{z}=\tilde{\protect\varepsilon}=0,$ $\protect\varepsilon %
_{x}=-0.1\protect\gamma $, $g$ (normalized to $\protect\gamma \protect\xi %
_{H}^{-2}$) equal to 40. There are switching between two types of the
solution: ($q>0,n>0$) and ($q=\max ,n=0$). The inset shows zoom of the
region near switching point. $T_{c}$ line consists of several curves each
corresponding to a different Landau level $n$ ($n=0,1$ and $2$ in this case).
}
\label{fig:Tcn}
\end{figure}

Using $q_{x}^{\prime }$ corresponding to the maximum of $\alpha (H,T)$ we
find $T_{c}(H)$. When the parameters of our system (the actual values of $%
\tilde{\varepsilon},\varepsilon _{z},\varepsilon _{x}$ coefficients)
correspond to the point in ($\tilde{\varepsilon},\varepsilon _{z}$) plane
situated near one of the separation lines in Figs.~\ref{fig:Hx} or \ref%
{fig:Hxy} then the magnetic field rotation can lead to transition between
the corresponding \bigskip two phases. The simplest case with $\tilde{%
\varepsilon}=0$, $\varepsilon _{z}=0$ is shown in the Figs.~\ref{fig:Tcp}
and \ref{fig:Tcn}. For positive $\varepsilon _{x}$ the transition is between
states $\left( q=0,n=\max \right) $ and ($q>0,n>0)$; for negative $%
\varepsilon _{x}$ the two states are $\left( q=\max ,n=0\right) $ and ($%
q>0,n>0)$. Integer nature of Landau level $n$ manifests itself in the state (%
$q>0,n>0)$, when the FFLO modulation could change several Landau levels
while magnetic field rotates in a given region. In this case $T_{c}$ line
consists of several curves each corresponding to a different Landau level
solution. If switching between the different solutions does not occur then
general $T_{cu}$ dependence will be reduced to the simple sinusoidal form
with period $\pi /4$.

\section{\protect\bigskip Cubic symmetry}

In the case of cubic symmetry $\varepsilon _{z}$ is equal to $0$ and $\tilde{%
\varepsilon}=\varepsilon _{x}=\varepsilon $. In the absence of the orbital
effect the direction of modulation will be along one of the axis if $%
\varepsilon >0$, and along one of the main diagonals if $\varepsilon <0$. In
the presence of orbital effect when magnetic field is applied along one of
the cubic axis the type of solution for maximum $\alpha (H,T)$ depends only
on the sign of $\varepsilon $. If $\varepsilon <0$ then $q$ is equal to $%
\sqrt{\frac{3}{7}u_{0}}$ and $n=\left[ \frac{2}{7}\xi _{H}^{2}u_{0}-\frac{1}{%
2}\right] $. For $\varepsilon >0$ there will be the choice between highest
and zero modulations. If $u_{0}>7\xi _{H}^{-2}$ then we obtain the maximum\
modulation along the field $q=\sqrt{u_{0}-\xi _{H}^{-2}}$ and Landau level $%
n $ will be zero. But when $u_{0}<7\xi _{H}^{-2}$ it is more favorable to
have zero (or some residual due to the mismatch) modulation $q$ and highest
Landau level $n=\left[ \frac{1}{2}\xi _{H}^{2}u_{0}-\frac{1}{2}\right] $. In
the latter case when the modulation along the field is absent ($q\approx 0$)
it turns out that the maximum Landau level can not be higher than 3. In the
case with magnetic field applied along the one of the diagonals we have the
same situation but with the opposite sign of $\varepsilon $. Note that in
all these cases the momentum and Landau level do not depend on $\varepsilon $
at the first approximation.

\section{Discussion}

\bigskip We investigated the influence of the crystal structure effects on
the FFLO state based on the modified Ginzburg-Landau approach. We analyzed
the possible solutions for the FFLO\ modulation vector and relevant Landau
level functions. We have used the single-level approximation, but we believe
that qualitatively our results would remain valid even if we take into
account the general multi-level representation of the order parameter. For
illustration we have restricted ourself to the tetragonal symmetry because
most promising material for FFLO realization CeCoIn$_{5\text{ }}$has namely
this type of the symmetry. Our results can be easily generalized to any
symmetry as long as deviation of the Fermi surface from the elliptic form
can be treated as a perturbation. In the opposite case the single-level
Landau function solution will be transformed into a series of higher level
functions. Also this will lead to the broadening of the $q=0$ region shown
in Fig.~\ref{fig:residual}, which means that for a wide range of parameters
in such a case there will be no more modulation along the field. The form of
the Fermi surface determines the direction of the FFLO modulation in the
pure paramagnetic limit. We see that in the presence of the orbital effect
the system tries in some way to reproduce this optimal directions of the
FFLO modulation by varying the Landau level index $n$ and wave-vector of the
modulation along the field.

The higher Landau level solutions has been predicted for the FFLO\ phase in
2D superconductors in tilted magnetic field \cite{BuzdinBrisonEuro,
Bulaevskii74, ShimaharaRainer97}, in 3D $d$-wave and quasi-2D $s$- and $d$%
-wave superconductors \cite{ShimaharaMatsuo96, ShimaharaSuginishi}, and in
3D isotropic superconductors at low temperature provided the Maki parameter
is large \cite{BuzdinBrison}. Here we have demonstrated that for certain
field orientations such states naturally appear in real 3D compounds in a
whole region of the FFLO phase existence (without any restriction to the
value of Maki parameter). This behavior is related with crystal structure
and/or pairing symmetry effects. The isotropic models used so far to
describe FFLO state fail to predict these different types of the scenarios
of the FFLO transitions. Indeed following the isotropic (or quasi-isotropic)
model the transition to the FFLO state with the increase of the magnetic
field always occurs via the modulation appearance along the field direction.
On the contrary in the present paper we predict the FFLO transition as a
formation of the higher Landau level states. The vortex state that
corresponds to these higher Landau level solutions have a rather complicated
structure due to the competition between two length scales, the average
distance between vortices and the FFLO period \cite{Houzet2000, Klein2000,
Yang2004}. \ Recently in \cite{Agterberg2008} the very special vortex phases
with spatial line nodes forming a variety of 3D spatial configurations has
been predicted. Therefore we may expect that the mixed state in the FFLO
superconductor may be very different from the usual Abrikosov lattice,
provided that the higher Landau level solutions are realized. The
experimentally verified consequences of these scenario of the FFLO
transition are the first order transitions between the states with different
Landau level solutions (namely between $n=0$ and $n=1$), accompanying by the
strong change of the vortex lattice structure. The standard experimental
techniques of the vortex lattice observation (including the neutron
scattering) could be used to detect these transformations.

\bigskip It is commonly believed that the FFLO state in CeCoIn$_{5\text{ }}$%
corresponds to the state with the modulation along the magnetic field for
both field orientations: along the tetragonal axis and in the basal plane.
However, comparing the ($\tilde{\varepsilon},\varepsilon _{z}$) diagrams
(Figs.~\ref{fig:z} and \ref{fig:Hx}) we see that the situation when we have
a zero Landau level solution for this two field orientations is improbable.
In CeCoIn$_{5\text{ }}$the crystal structure effects are rather important --
for example in \cite{DeBeer06} the vortex lattice reorientation transition\
have been reported as well as in-plane anisotropy of the upper critical
field \cite{Weickert06}. In such a case we can expect that for one of these
field orientations the Landau level solution with $n\geq 1$ may be realized.
Note that such a possibility in connection with the FFLO state in CeCoIn$_{5%
\text{ }}$has been discussed in \cite{Radovan03}. If the crystal structure
effects are large enough for the Landau level solutions with $n\geq 1$ the
modulation along the field may be absent. Very recently \cite{Kenzelamann}
the modulated antiferromagnetic ordering has been reported in the low
temperature superconducting phase of CeCoIn$_{5\text{ }}$at the magnetic
field in the basal plane. The antiferromagnetic ordering plays for the FFLO
state the role of the crystal structure effect favoring the orientation of
the FFLO modulation wavevector along the antiferromagnetic one \cite%
{Brison95}. The texture in the superconducting order parameter revealed by
NMR\ experiments looks different for different field orientations \cite%
{MatsudaShimahara07} as well as the anomaly in the local magnetic inductor
measurements \cite{Okazaki07}. This may indicate on the different types of
the FFLO state for different field orientations. Presumably for the field
orientation in the basal plane there are no FFLO modulation along the field.

\begin{acknowledgments}
The authors are grateful to Y. Matsuda, T. Shibauchi, J.P. Brison, J.
Flouquet, J. Cayssol and F. Konschelle for useful discussions and comments.
This work was supported by ANR Extreme Conditions Correlated Electrons
ANR-06-BLAN-0220 (for Dmitry Denisov and Alexander Buzdin) and by University
Bordeaux 1 (for Hiroshi Shimahra).
\end{acknowledgments}

\end{document}